# Theoretical Study of Coulomb Correlations and Spin-Orbit Coupling in SrIrO$_3$

Vijeta Singh[*] and J. J. Pulikkotil

*National Physical Laboratory, Council of Scientific and Industrial Research, New Delhi-110012, India*
[*]*E- mail: vijetasingh@nplindia.org*

**Abstract.** Given that energy scales associated with crystal field splitting, spin orbit coupling and coulomb correlations in iridates are comparable, hence leading to exotic properties, we investigate the physical properties of orthorhombic SrIrO$_3$ using density functional theory. Our calculations, however, show that SrIrO$_3$ is a bad metal with no long range magnetic ordering, unlike its sister compounds Sr$_2$IrO$_4$ and Sr$_3$Ir$_2$O$_7$. Moreover, despite having large band width, it appears conclusive that the larger resistivity in SrIrO$_3$ is due to spin orbit interactions. Besides, the effects of electron-electron correlations on its electronic structure and magnetic properties are also discussed.

## INTRODUCTION

Iridates are expected to have a conducting ground state, due to their larger spatial extent of the 5d orbitals and hence higher orbital overlap with its near neighbours. However, the case appears contradictory. Experimental reports on a variety of oxides containing iridium show that they are either weakly conducting or insulating, thereby posing a new quantum paradigm in solid state. The Ruddlesden-Popper series of iridate perovskites with generic formula Sr$_{n+1}$Ir$_n$O$_{3n+1}$ warrants special attention, in this regard [4, 5]. Both Sr$_2$IrO$_4$ (n = 1) and Sr$_3$Ir$_2$O$_7$ (n = 2) manifest a new quantum phase, otherwise known as "spin-orbit driven J$_{eff}$ = $1/2$ Mottt insulators". In these systems, the octahedral IrO$_6$ motif facilitate a t$_{2g}$ - e$_g$ crystal field splitting of the 5d- bands, which under the influence of spin-orbit interactions are further split into a four-fold degenerate J$_{eff}$ = $3/2$ quadruplet and a two-fold J$_{eff}$ = $3/2$ doublet states. While the former states are lower in energy and is completely filled, the valence count of the Ir$^{4+}$ ions (with 5d$^5$ electronic configuration) attribute a single electron occupancy to the J$_{eff}$ = $1/2$ doublet staste, which further split into an upper and lower Hubbard band leading to an insulating ground state [2, 5]. Thus, it become quite evident in iridates that the energy length-scales associated with the crystal field, Coulomb correlations and spin-orbit interactions are comparable and exotic properties such as those of topological insulators, topological Mott insulators and Weyl semi-metals may arise [3, 8].

However, it also seems that the underlying structural dimensionality also plays an important role. While the n = 1 and 2 compounds of Sr$_{n+1}$Ir$_n$O$_{3n+1}$ series crystallize in a quasi-two dimensional structures, the case referring to n = ∞ i.e., SrIrO$_3$ is three dimensional and are either semi-metallic or metallic, with subtle signature of electron correlations [1]. The nature of magnetic interactions also change with structure for example, Sr$_2$IrO$_4$ and Sr$_3$Ir$_2$O$_7$ are anti-ferromagnetic, while both polymorphs of SrIrO$_3$, monoclinic and orthorhombic, has no long range magnetic ordering deep down to low temperatures. The electronic and magnetic structure of orthorhombic phase, however, remains less explored in detail and therefore forms the objective of this work. The results presented are based on the all-electron full-potential linearized augmented plane-wave (FP-LAPW) method, as implemented in the WIEN2K suite of programs [6]. The lattice parameters with a = 5.591Å b = 7.882Å, and c = 5.561Å, following Ref. [7] were adopted. The exchange-correlation potential was described in local density approximation (LDA) and generalized gradient approximation (GGA). To check for the effects of Coulomb correlations in $SrIrO_3$, we also employed the GGA+$U_{eff}$ scheme with $U_{eff}$ varied from 0 to 4 eV.

## RESULTS AND DISCUSSIONS

### A. Structural Optimization

The larger extent of the Ir 5d orbitals in $SrIrO_3$ suggest a stronger interaction between the 5d and O 2p, implying a tendency to form distorted structures. This structural effect, in turn, could narrow the 5d band width bringing the system close to an insulating phase. The theoretical refinements of the $SrIrO_3$ structure deduce that the Ir−O(1)−Ir and Ir−O(2)−Ir bond angles, (*O(1)* and *O(2)* being the in-plane and out-of-plane oxygen ions) of the Pnma symmetry, are 152.39° and 153.35°, respectively and that the Ir−O(1) and Ir−O(2) bond distances are 2.029 and 2.025 Å, respectively. The calculated values are in reasonable agreement with the experiments [9].

### B. Electronic Structure

In Fig. 1(a) we show the GGA generated non-magnetic density of states (DOS) of $SrIrO_3$. The Ir 5d and the O 2p DOS are spread over the entire valence band energy, suggesting strong covalent hybridization between them. A $t_{2g}$ - $e_g$ crystal field splitting of the Ir 5d bands, by $\simeq$ 0.6 eV is evident in the DOS spectra. The states over the range $-8 \leq E(eV) \leq +0.5$ represents the $t_{2g}$ states, while those above $E \geq +1.1$ eV constitute the $e_g$ states. However, due to the distortion in the $IrO_6$ octahedra, the $t_{2g}$ motif are further split, represented by a singularity like feature at −2.2 eV below the Fermi energy. The calculated density of state at Fermi energy is 2.11St./eV.

In Fig. 1 (b) we show the modulations in the DOS spectra of $SrIrO_3$ when subjected to spin orbit coupling. While the bottom of the valence states over the energy range $-8 \leq E(eV) \leq -1$, remains more or less invariant, the Ir 5d states exhibit dramatic changes in the vicinity of the Fermi energy. The most interesting feature in the spectra is the sharp reduction in the DOS at Fermi energy. This value (1.16St./eV) decreases by $\simeq$ 45% in comparison to scalar-relativistic calculation. The main consequence of spin-orbit effects is the separation of the $J_{eff}$ states into a four-fold degenerate $J_{eff} = 3/2$ quadruplet and a two-fold $J_{eff} = 1/2$ doublet states. However, these states are strongly hybridized, thus yielding a finite DOS at the Fermi energy.

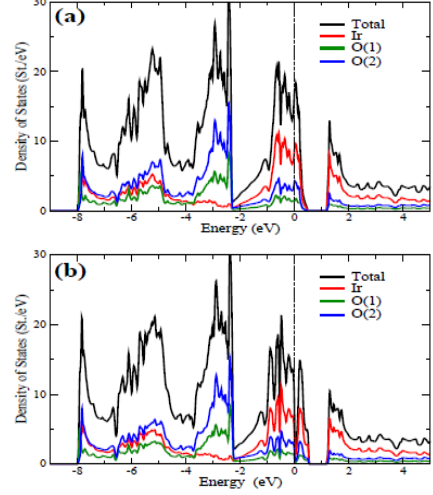

Figure1: (Color) The GGA generated non-magnetic DOS of $SrIrO_3$ calculated using the scalar relativistic Hamiltonian with (a) no spin-orbit coupling and, (b) with spin-orbit coupling treated in the second variational approach. The vertical line through energy zero represents the reference Fermi energy.

To understand the effects of Coulomb correlations in $SrIrO_3$, we adopt the GGA+$U_{eff}$ formalism. For $U_{eff} <$ 4 eV, we find $SrIrO_3$ to have a metallic ground state, however, with small but finite DOS at the Fermi energy. When $U_{eff} >$ 4 eV, the system becomes an insulator. Although, an exact determination of $U_{eff}$ is at presently beyond the scope of this work, we find that an insulating ground state emerges only when $U_{eff} >$ W, where W is the band width of Ir 5d orbitals. The striking difference between the crystal structures of $Sr_2IrO_4$ and $Sr_3Ir_2O_7$ and that of $SrIrO_3$ is that while the former are quasi-two dimensional, the latter adopts to a 3-D structure. Thus, beyond the factors, such as electron-electron interactions and spin-orbit coupling that render an insulating phase in iridates, we identify that dimensionality also plays an important role in these class of materials.

### C. Magnetic Properties

We first check the propensity of ferromagnetic ordering in SrIrO$_3$ using the Stoner criteria, which suggests that a system would undergo FM splitting of states, provided it satisfies the inequality N(E$_F$)*I ≥1, where N(E$_F$) is the non-magnetic DOS at Fermi energy, and I the Stoner parameter. The I for Ir is calculated as 0.574.according to the Ref. [11]. This suggests that SrIrO$_3$ with spin-orbit term included in the Hamiltonian render a non-magnetic structure, since the Stoner product yields 0.563. However, in the scalar relativistic formalism, the product becomes 1.2, inferring to a ferromagnetic ordering of Ir ions in SrIrO$_3$.

Both in LDA and GGA, SrIrO$_3$ is metallic in all studied magnetic structures. In the conventional LDA approach we find no magnetic solutions, which is in contrast to earlier reports [10]. On the other hand, GGA calculations yield a finite ferromagnetic moment of 0.13μ$_B$ at the Ir sites. However, with inclusion of spin-orbit term becomes the crystal Hamiltonian, the ground state becomes non-magnetic. Besides, no magnetic solutions were obtained for the AFM.

| without spin-orbit | | | | |
|---|---|---|---|---|
| U$_{eff}$ | FM | AFM-A | AFM-C | AFM-G |
| 0 | 0.13 | × | × | × |
| 2 | 0.56 | 0.44 | × | 0.20 |
| 4 | 0.58 | 0.58 | 0.52 | 0.50 |
| With spin-orbit | | | | |
| 0 | × | × | × | × |
| 2 | × | × | × | × |
| 4 | 0.53 | 0.53 | 0.48 | 0.47 |

Table I: The GGA+U$_{eff}$ (U$_{eff}$ in eV) computed magnetic moment (μ$_B$), with (in parentheses) and without spin-orbit interactions.

In Table I, we show the computed values of Ir magnetic moment with U$_{eff}$. These calculations show that for U$_{eff}$ < W, the system is itinerant, i.e., the magnitude of the Ir moment depends strongly on the local magnetic environment. While for G-AFM structure in the scalar relativistic mode, while the moment is determined to be 0.2 μ$_B$, it is almost twice larger for FM and A-AFM structures. However, with inclusion of spin-orbit, the moment vanishes. In case, for U$_{eff}$ > W, the nature of magnetism transforms to a localized picture, i.e., the magnitude of the moment remains invariant (≃ 0.5μ$_B$), irrespective of the local magnetic neighbourhood. These values, remain constant, irrespective of the mode of calculation i.e., with and without spin-orbit interactions. Thus, the proper estimation of U$_{eff}$ in SrIrO$_3$ is crucial, especially when it comes to study the excitation modes. However, the lack of photo-emission spectra leaves us at this point to only qualitatively infer the effects of U$_{eff}$ on the electron structure of SrIrO$_3$.

## SUMMARY AND CONCLUSION

Our calculations on SrIrO$_3$ suggest that both spin-orbit coupling and electron-electron correlations are weak in this system. The magnetic phase diagram of SrIrO$_3$ in the GGA+U$_{eff}$ approach, finds that for small U$_{eff}$, (U$_{eff}$ < W) no long range magnetic ordering is feasible, while for U$_{eff}$ >W, anti-ferromagnetic interactions appear dominant.

## ACKNOWLEDGMENTS

Financial assistance from CSIR XII FYP project AQuaRIUS is gratefully acknowledged. Vijeta Singh acknowledges financial support from CSIR India.